\documentclass[aps,twocolumn]{revtex4}
\usepackage[dvips]{graphicx}
\usepackage{latexsym}
\usepackage{natbib}
\usepackage{subfigure}
\usepackage{color}

\begin{document}

\title{ Cascading Failures in Interdependent Lattice Networks:\\
The Critical Role of the Length of Dependency Links}
\author{ Wei Li$^1$, Amir Bashan$^2$, Sergey V. Buldyrev$^{3}$,
H. Eugene Stanley$^1$, and Shlomo Havlin$^1$$^2$}
\affiliation{$^1$Center for Polymer Studies and Department of Physics,
Boston University, Boston, MA 02215 USA\\
$^2$Department of Physics, Bar-Ilan University, Ramat-Gan 52900, Israel\\
$^3$Department of Physics, Yeshiva University, 500 West 185th Street, New
York, New York 10033, USA}
\date{1 June 2012 }

\begin{abstract}

We study the cascading failures in a system composed of two
interdependent square lattice networks A and B placed on the same
Cartesian plane, where each node in network A depends on a node in
network B randomly chosen within a certain distance $r$ from the
corresponding node in network A and vice versa. Our results suggest
that percolation for small $r$ below $r_{\rm max}\approx 8$ (lattice units)
is a second-order transition, and for
larger $r$ is a first-order transition.  For $r<r_{\rm max}$, the
critical threshold increases linearly with $r$ from 0.593 at $r=0$ and
reaches a maximum, $0.738$ for $r=r_{\rm max}$ and then gradually
decreases to 0.683 for $r=\infty$. Our analytical considerations are in
good agreement with simulations. Our study suggests that interdependent
infrastructures embedded in Euclidean space become most vulnerable when
the distance between interdependent nodes is in the intermediate range,
which is much smaller than the size of the system.

\end{abstract}

\maketitle

Most previous studies of the robustness of interdependent networks
\cite{Buldyrev2010,Alessandro2010,Parshani2010prl,Parshani2010epl,Shao2011,Leicht2011,D'Souza2011,Xu2011,Hao2011,Kai2011,Gu2011,Huang2011,Gao2011,Hu2011,Sergey2011pre,Parshani2011,Bashan2011,Bashan2011pre,Gao2010arxiv}
focused on random networks in which space restrictions are not
considered. Most real networks are embedded either in two-dimensional or
in three-dimensional space, and the nodes in each network might be
interdependent with nodes in other networks. One example is a computer
in a computer network is dependent on power upon the functioning of a
local power grid network where both networks are spatially
embedded. Another example is the way the world-wide network of seaports
embedded in the two-dimensional surface of the earth is interdependent
with power grid networks embedded on the same surface. A seaport needs
electricity from a nearby power station to operate and a power station
needs fuel supplied through a nearby seaport to operate. Thus the
failure of a power station in a power grid network will cause a failure
in a nearby seaport and vice versa.  Space constraints, such as the
network dimensionality \cite{Li2011dimension}, influence the network
properties dramatically, and thus the question about the resilience of
interdependent spatial networks is of much interest.

\begin{figure}
\includegraphics[width=0.4\textwidth, angle = 0]{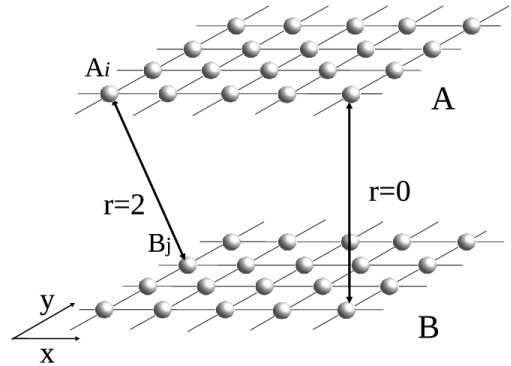}
\caption{Two square lattices A and B where in each lattice every node
  has two types of links: connectivity links and dependency links.
  Every node is initially connected to its four nearest neighbors within
  the same lattice via connectivity links.  Also, each node $A_i$ in
  lattice A depends on one and only one node $B_j$ in lattice B via a
  dependency link (and vice versa), with the only constraint that
  $|x_i-x_j|\leq r$ and $|y_i-y_j|\leq r$. If node $A_i$ fails, then
  node $B_j$ fails. If node $B_j$ fails, then node $A_i$ fails. Network
  A is shifted vertically for clarity.
\label{model.ps}}
\end{figure}

  The case of interdependent spatially embedded networks
  is significantly different from interdependent random networks in two
  ways: (i) within each network, nodes are connected only to the nodes
  in their spatial vicinity, while in the randomly connected networks,
  the concept of spatial vicinity is not defined; (ii) the dependency
  links establishing the interdependence between the networks might not
  be random but may have a typical length $r$. To understand how these space
constraints affect the resiliency of interdependent networks, we study
the mutual percolation of a system composed of two interdependent
two-dimensional lattices A and B, where a node $A_i$ can
  connect to its dependent node $B_j$ only within distance $r$ from
  $A_i$ (see Fig.~\ref{model.ps}). Since a node can be functional only
  if it is connected to the network, the resilience can be measured,
  using percolation theory, as the size of the remaining giant component
  after an attack on network.

Our model consists of two identical square lattices A and B of linear
size $L$ and $N=L^2$ nodes with periodic boundary conditions. In each
lattice, each node has two types of links: connectivity links and
dependency links. Each node is connected to its four nearest neighbors
within the same lattice via connectivity links.  Also, a node $A_i$
located at $(x_i,y_i)$ in lattice A is connected with one and only one
node $B_j$ located at $(x_j,y_j)$ in lattice B via a bidirectional 
dependency link,
with the only constraint that $|x_i-x_j|\leq r$ and $|y_i-y_j|\leq r$
(Fig.~\ref{model.ps}). The parameter $r$ represents the maximum
distance a node in one network gets support from a node in another
network.

Although real networks embedded in two-dimensional space may have more
complex structures than the square lattice, our model can serve as a
benchmark for more complex situations.  Moreover, it is known that the
percolation transition in two dimensions has universal scaling behavior
which does not depend on the coordination number and is the same for
lattice and off-lattice models, as long as the links
  have a finite characteristic length. Hence mutual percolation in two
dimensions should not depend on the particular realization of the model \cite{Bunde1991}.

\begin{figure}
\includegraphics[width=0.4\textwidth, angle = -90]{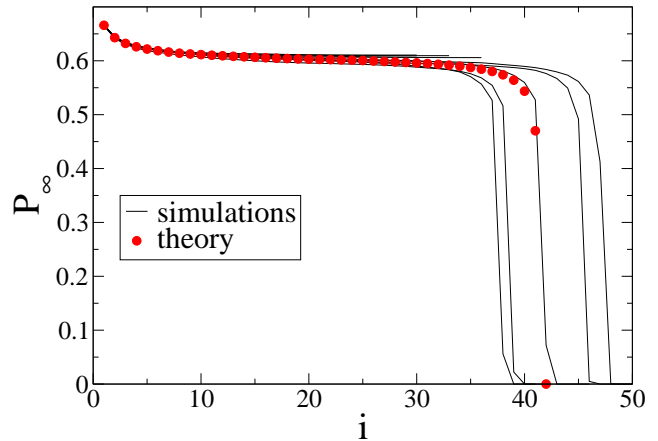}
\caption{Giant component size $P_\infty$ as a function of step $i$ at
  the first-order transition regime at $p=0.6825$ for $r=L=1000$.  The
  simulation results (solid lines) are in good agreement with the
  theoretical results (dots). The value of $p$ is close to the
  percolation threshold $p_c^\mu=0.6827$.
\label{iter.ps}}
\end{figure}

\begin{figure}
\includegraphics[width=0.4\textwidth, angle = -90]{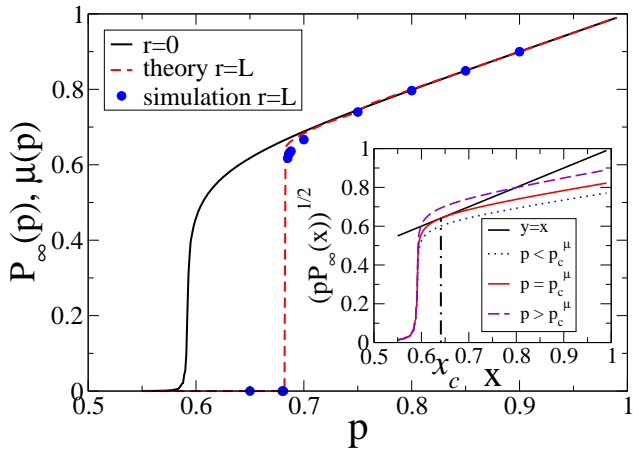}
\caption{The giant component size $P_\infty$ as a function of remaining
  fraction of nodes $p$. The solid curve is for conventional percolation
  on a single square lattice, which also describes the
    limiting case of $r=0$.  The solid curve is obtained by numerical
  simulations on $N=4000\times4000$ lattice sites with periodic boundary
  conditions and averaged over 100 realizations. The dash curve
  represents the theoretical result $\mu(p)$ for two interdependent lattice
  networks with $r=L$ given by Eq.~(\ref{e:pinf}). The simulation
  results (dots) are for two interdependent lattice networks with
  $N=1000\times1000$ and $r=L$.  Inset: A schematic graphical solution
  of Eq.~(\ref{e:pinf}) is shown.  The curves are $\sqrt{pP_\infty(x)}$
  for different $p$ and the solution of Eq. (\ref{e:pinf}) is given by
  the intersection of the solid curves and the straight line $y=x$. The
  critical $p=p_c^\mu$ corresponds to the case when the solid curve is
  tangential to the straight line $y=x$.  Numerical solutions of
  Eqs.~(\ref{e:pc1}) and (\ref{e:pc2}) yield $x_c=0.641$,
  $P_\infty(x_c)=0.602$, and $p_c^\mu=0.6827$.
\label{infp.ps}}
\end{figure}

The difference between connectivity and dependency links is that for
connectivity links, a node fails only when it does not belong to the
giant cluster of its network, while for dependency links, a node fails
once the node on which it depends in the other network (connected via a
dependency link) fails. An initial random attack destroys a fraction
$1-p$ of nodes in network A. This causes a certain number of nodes to
disconnect from the giant component of network A so that only a fraction
of nodes $p_1=P_\infty(p)$ remains functional. Here $P_\infty(p)$ is the
order parameter of conventional percolation in a square lattice
\cite{Bunde1991}. The removal nodes in network A
causes the removal of the dependent nodes in network B. As a result,
only a fraction $P_\infty(p_1)$ of nodes in network B remains
functional. This produces additional damage in network A and so on. The
cascading failure process stops when no further damage propagates
between the lattices.  If the length of dependency links is totally
random ($r=L$), the formalism developed in Ref.~\cite{Buldyrev2010} can
be applied. This is because at the $i$-th stage of the cascade the resulting
giant component $P_\infty(p_i)$ is the order parameter of conventional
percolation computed for a random fraction of nodes $p_i$ surviving
after all the nodes in one network that depend on the nonfunctional nodes
of the other network are removed. Accordingly we can represent the
cascading failure by the recursive equations for the survived fraction
$p_i$,
\begin{eqnarray*}
\\&p_0=p,
\\&p_1=\frac{p}{p_0}P_\infty(p_0)=P_\infty(p),
\\&\vdots
\end{eqnarray*}
\begin{equation}
p_i=\frac{p}{p_{i-1}}P_\infty(p_{i-1}).
\label{e:rec}
\end{equation}
The recursive steps of Eq.~(\ref{e:rec}), representing the cascading
failures in the giant component shown in Fig.~\ref{iter.ps}, are in good
agreement with simulations. In the limit $i\to\infty$,
{Eq.}~(\ref{e:rec}) yields the equation for the mutual giant component
at steady state, $\mu\equiv P_\infty(p_\infty)$,
\begin{equation}
x=\sqrt{pP_\infty(x)},
\label{e:pinf}
\end{equation}
where $x\equiv p_\infty$. Using the form of $P_\infty(x)$ for
conventional percolation obtained from numerical simulations,
Eq.~(\ref{e:pinf}) can be solved graphically as shown in the inset of
Fig.~\ref{infp.ps}.  Due to the specific shape of the function
$P_\infty(p)$ [see Fig.~\ref{infp.ps}], ($P_\infty(p)<p$, $\lim_{p\to
  1}P_\infty/p=1$, $\lim_{p\to p_c} P_\infty(p)= 0$, and $p_c=0.5927$
for square lattice), it does not have solutions for a small $p$ except
for the trivial case $x=0$.

Figure~\ref{infp.ps} shows the numerical solution of Eq.~(\ref{e:pinf})
which is in good agreement with simulations and compares it with
$P_\infty(p)$ of a single network. The critical $p$ for which the
nontrivial solution ceases to exist, $p\equiv p_c^\mu$, corresponds to
the case when the r.h.s. of Eq.~(\ref{e:pinf}) becomes tangential at the
point of their intersection $x=x_c$ to its l.h.s. (Fig.~\ref{infp.ps}
inset).  Hence
\begin{equation}
P'_\infty(x_c) x_c=2 P_\infty(x_c),
\label{e:pc1}
\end{equation}
from which the critical $p$ for mutual percolation is
\begin{equation}
p_c^\mu= x_c^2/P_\infty(x_c).
\label{e:pc2}
\end{equation}
Numerical solutions of Eqs.~(\ref{e:pc1}) and (\ref{e:pc2}) yield
$p_c^\mu=0.683$, $x_c=0.641$, and $P_\infty(x_c)=0.602$, in good
agreement with simulations of the mutual percolation on lattices
for $r=L$ as seen in Fig.~\ref{infp.ps}. Fig.~\ref{infp.ps} shows 
a discontinuity in the order parameter of
mutual percolation $\mu(p)=P_\infty(p)$ at $p=p_c^\mu$, which drops from
$\mu(p)=0.602$ to zero for $p<p_c^\mu$, characteristic of a first-order
transition.

\begin{figure}
\centering
\subfigure[]{\includegraphics[width=0.2\textwidth, angle = -0]{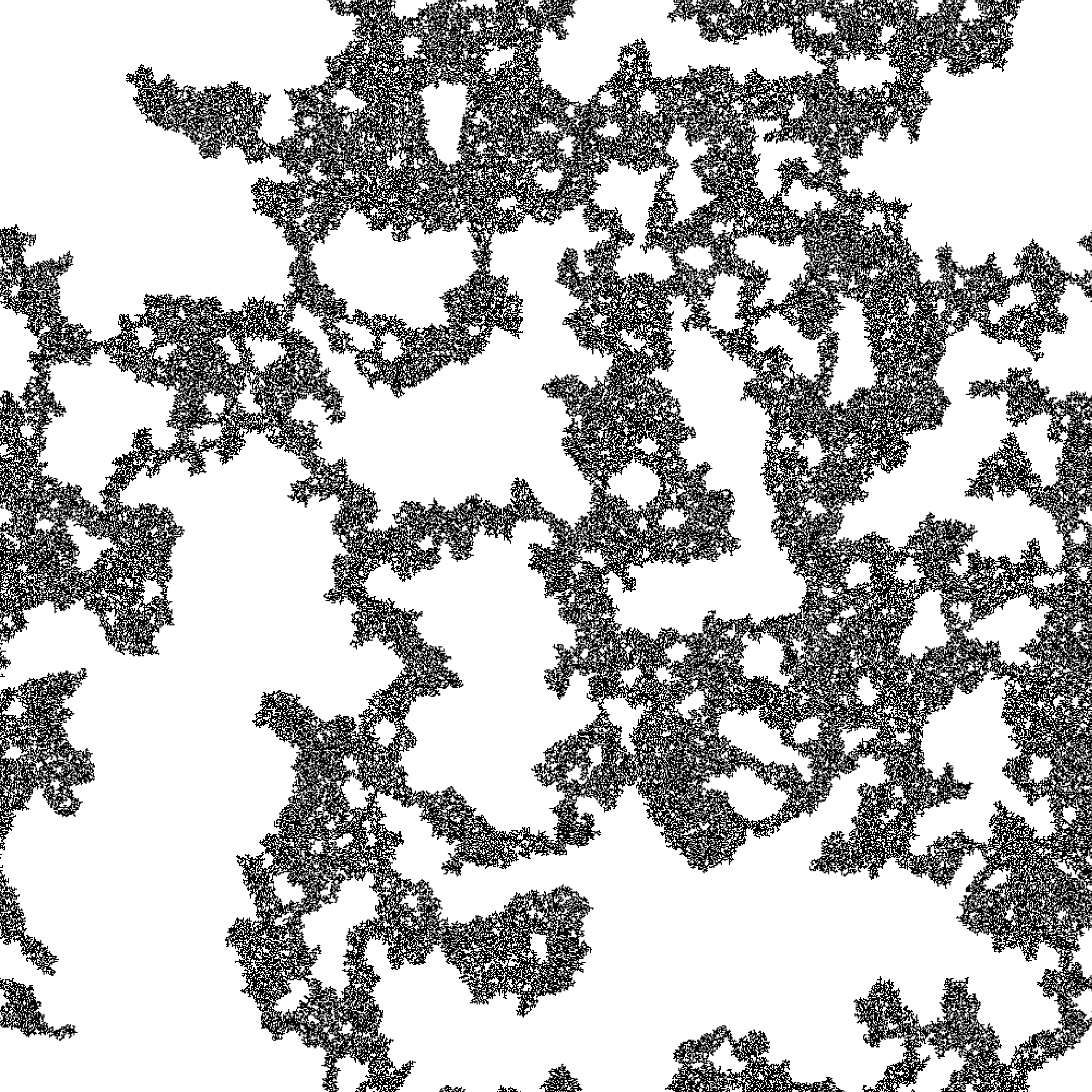}}
\subfigure[]{\includegraphics[width=0.2\textwidth, angle = -0]{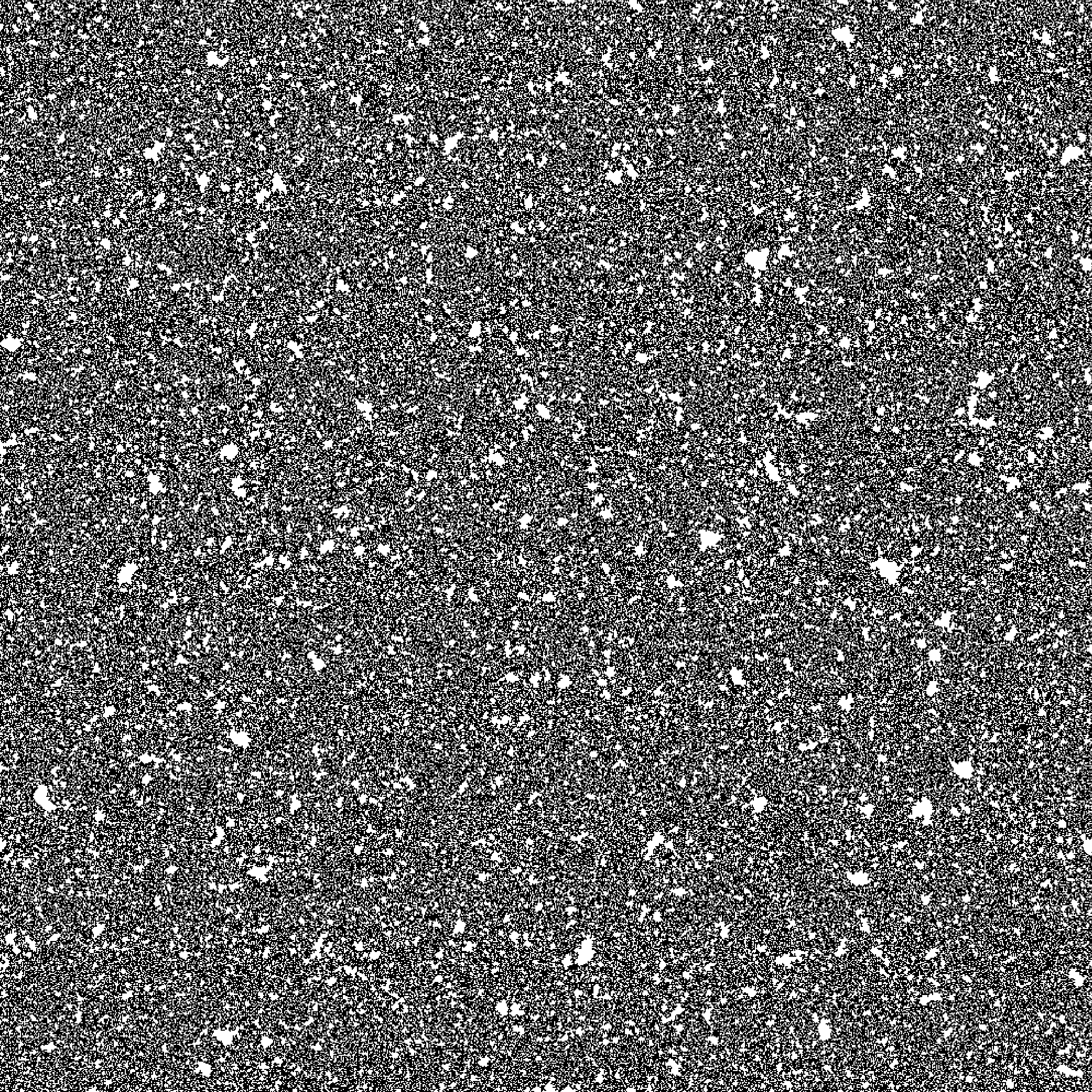}}
\subfigure[]{\includegraphics[width=0.15\textwidth, angle = -0]{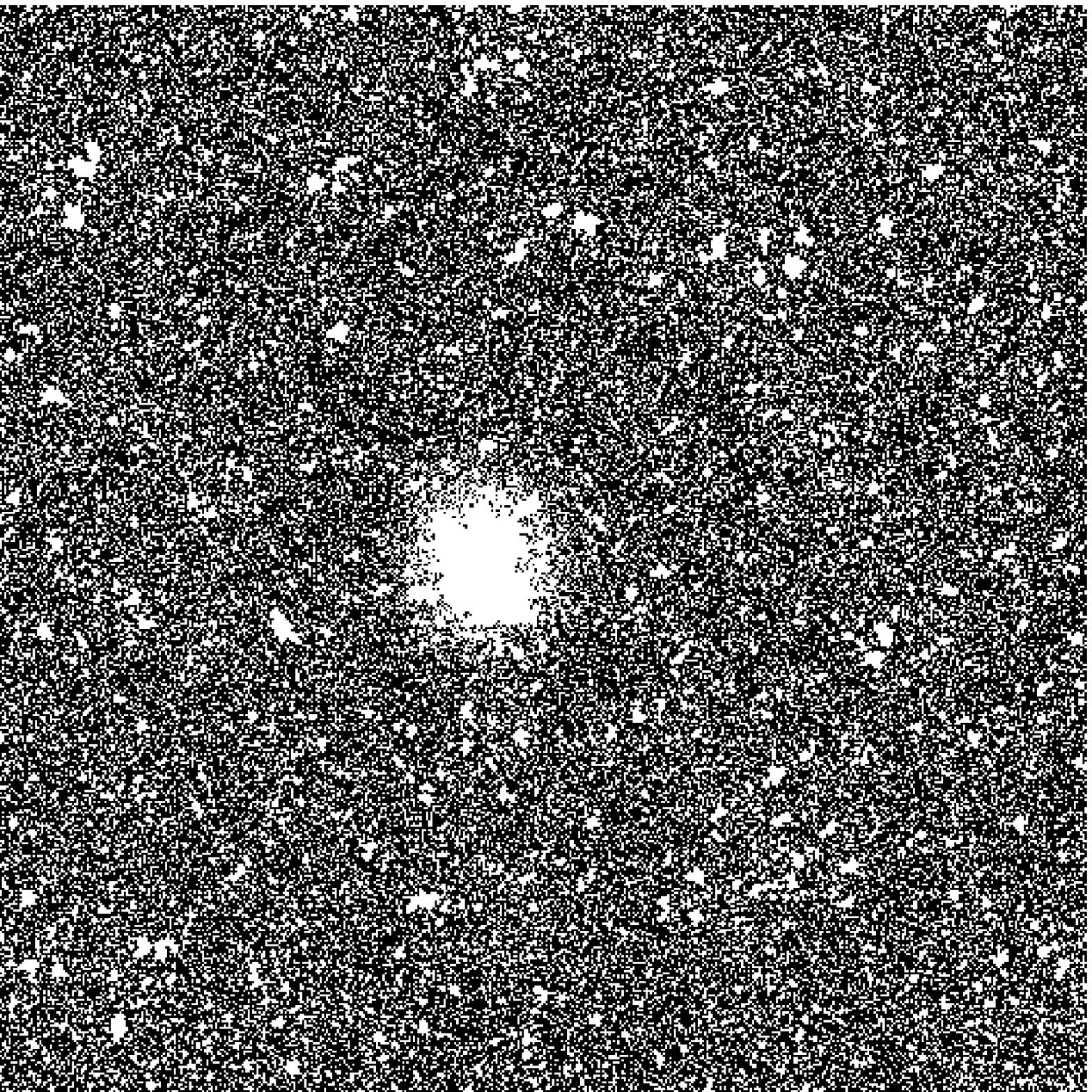}}
\subfigure[]{\includegraphics[width=0.15\textwidth, angle = -0]{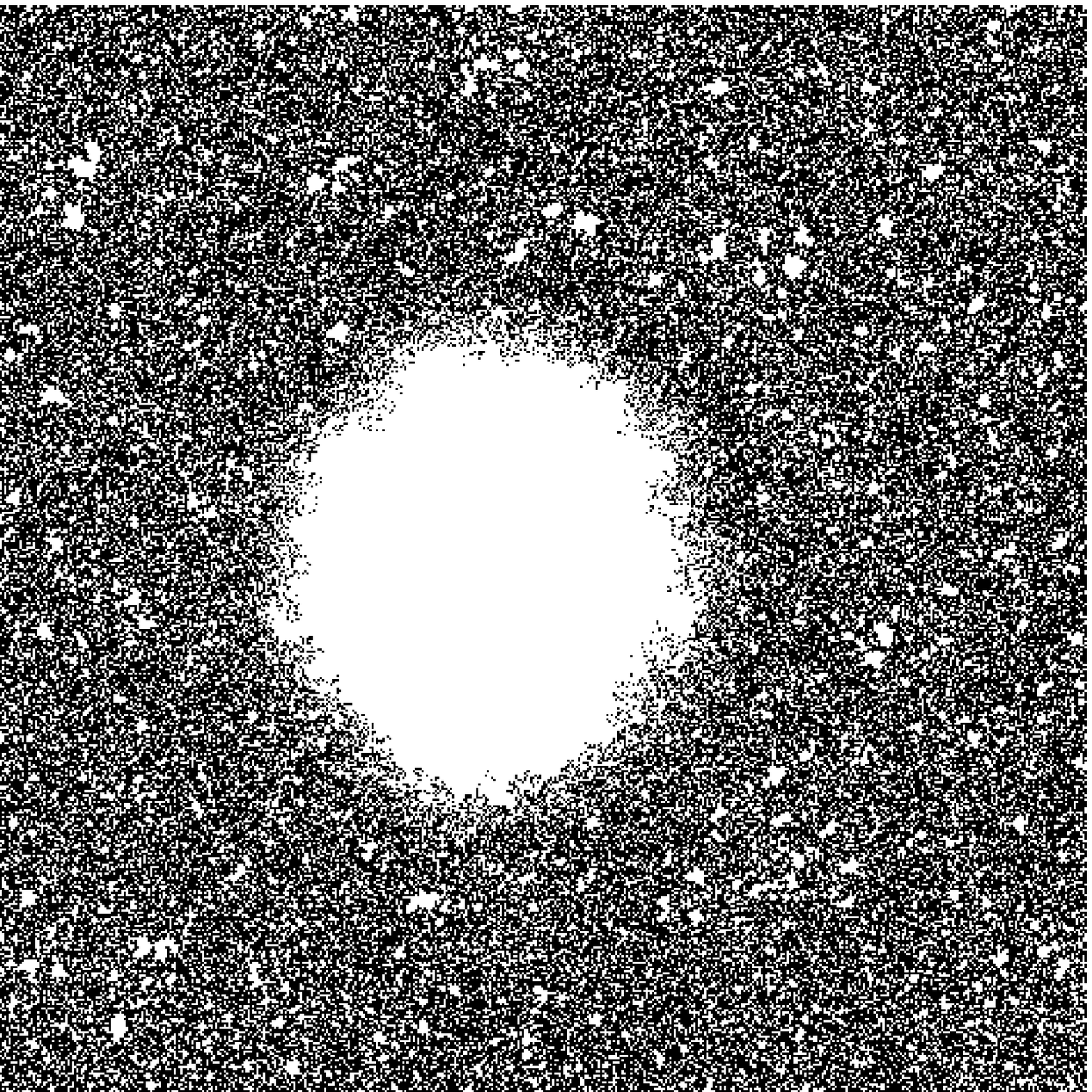}}
\subfigure[]{\includegraphics[width=0.15\textwidth, angle = -0]{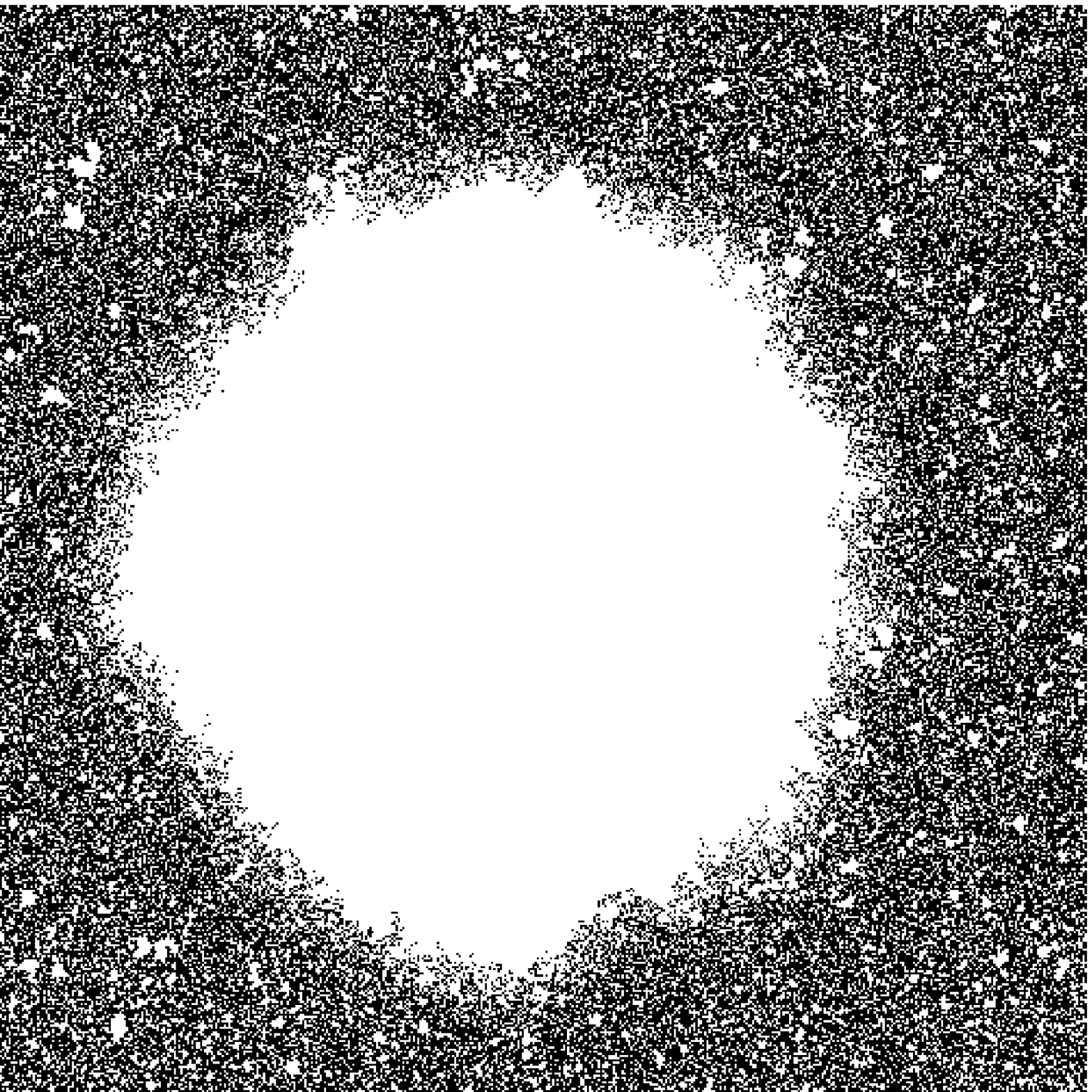}}
\caption{Three different typical behaviors of interdependent lattices
  near criticality. Pictures of stable mutual giant component at
  criticality of two interdependent lattices ($N=1000\times1000$) after
  cascading failures initiated by a random removal of $1-p$ of the nodes
  for (a) $r=4$ and $p=0.680$ and for (b) $r=1000$ and $p=0.683$. The
  dynamics of a growing bubble (explained in the text) for $r=20$ is
  demonstrated by three snapshots, (c), (d) and (e), of the non-stable
  giant component of the interdependent lattices ($N=500\times500$)
  during the cascading process initiated with $p=0.700$.}
\label{pictures}
\end{figure}

  Next, we study the mutual percolation for different
  dependency lengths $r$.  An infinite coupling distance $r=\infty$
  corresponds to the scenario of random dependency links between the
  lattices discussed above.  For $r=0$, every failed node in network A
  leads to removal of a node in network B in the same location. Thus,
  the percolation clusters in the two lattices are identical and there
  is no feedback failure in network A.  Therefore, the case of $r=0$ is
  identical to the case of conventional percolation in non-coupled
  lattices. Figures~\ref{pictures}(a),~\ref{pictures}(b) show the
structure of the giant component just above $p_c^\mu$ for very small $r$
(few lattice units) and for $r=L$ respectively. For small $r$ the
structure is similar to the heterogeneous fractal-like giant component
of a single network \cite{Bunde1991}. In contrast for $r$ of the order of
$L$ the giant component is homogeneous and almost 
compact [see Fig.~\ref{pictures}(b)] but,
surprisingly, on the verge of a sudden collapse as a first-order
transition. For intermediate values of $r$ the collapse occurs in a very
different way.
Figures~\ref{pictures}(c)--~\ref{pictures}(e) show for
intermediate values of $r$ (discussed below) that the initial cascade of
failures is localized to a region of size $r$ [Fig.~\ref{pictures}(c)]. Because of local density
fluctuations, the effective fraction of nodes $p$ in one region can be
smaller than the overall average, and therefore small clusters at this
region become isolated from the giant component and fail even when the
entire lattice is still connected. As soon as a region of size $r$
fails, the system becomes unstable: the interface of this bubble starts
to expand and soon engulfs the entire system
[Fig.~\ref{pictures}(d) and ~\ref{pictures}(e)]. This local effect of a propagating
interface owing to finite dependency links increases the system
vulnerability compared to the case of random dependency links.  Thus we
expect, $p_c^\mu(r)>p_c^\mu(\infty)$ found for random dependency
links. The process of formation of the critical bubble is similar to
nucleation near the gas-liquid spinodal \cite{Glaser1952}.  Thus, it is
important to understand the propagation of a flat interface, which would
correspond to gas-liquid coexistence.

In order to systematically study the conditions for propagation of a
flat interface, we study the two interdependent networks with an empty
gap on one edge in lattice A. We construct the two networks with the
length of interdependent links less than or equal to $r$ (see
Fig.~\ref{model.ps}).  The only difference from our original system is
that after random removal of a certain fraction of nodes $1-p$, we
eliminate the nodes in lattice A with coordinates distance $y_i\leq r$
to create an artificial flat interface. Simulations show that the flat
interface freely propagates and that the interdependent lattices system
totally collapses if $p<p_c^f(r)$, where $p_c^f(r)$ is approximately a
linear function of $r$ with $p_c^f(0)=p_c=0.5927$, $p_c^f(r_f)=1$, and
$r_f\cong 38$. For $r>r_f$, the interface freely propagates through the
system even when the lattice is completely intact. This happens because
the removed nodes of lattice A above the interface eliminate half of the
nodes in lattice B with $y_j\leq r$.  Thus the effective concentration
of nodes in lattice B linearly changes from $p$ at distance $r$ from
the interface to $p/2$ right at the interface. This system is analogous
to percolation in diffusion fronts studied by Sapoval et
al. \cite{Sapoval1985}. There is thus a certain distance from the
interface $r_c=r(2p_c-p)/p$ that corresponds to the critical threshold
of conventional percolation. If $r_c$ is much larger than the typical
cluster size in the range between $p_c$ and $p/2$, all the nodes in
lattice B in this layer will be disconnected and hence the interface
will propagate freely. The interface can stop if $r_c=\xi(p/2)$, {\it
  i.e.}, the connectedness correlation length \cite{Bunde1991} when
${p/2}$ is less than $p_c$. We estimate the critical concentration
$p_c^f$ from the equation $\xi(p_c^f/2)=r(2p_c-p_c^f)/p_c^f$, which
yields $r_f=\xi(1/2)/(2p_c-1)=41$ for the case $p=1$, where
$\xi(1/2)=7.6$ obtained by numerical simulations of conventional
percolation on a single lattice.  This prediction agrees well with
simulations ($r_f\cong 38$).  The propagation of the flat interface
close to $p_c^f(r)$ is similar to invasion percolation, which is a
fractal process with vanishing number of active sites, and the average
interface velocity approaches zero at $p_c^f(r)$, a characteristic of a
second-order transition. Thus, the system completely collapses when (1)
a flat interface exists and (2) $p<p_c^f$. The conditions for flat
interface propagation, $p_c^f(r)$ were obtained for the artificial model
where the flat interface is initially created. However, when the system
is initiated by a spatially random removal, a flat interface may be
created by random fluctuations over the lattice.

\begin{figure}
\includegraphics[width=0.4\textwidth, angle = -90]{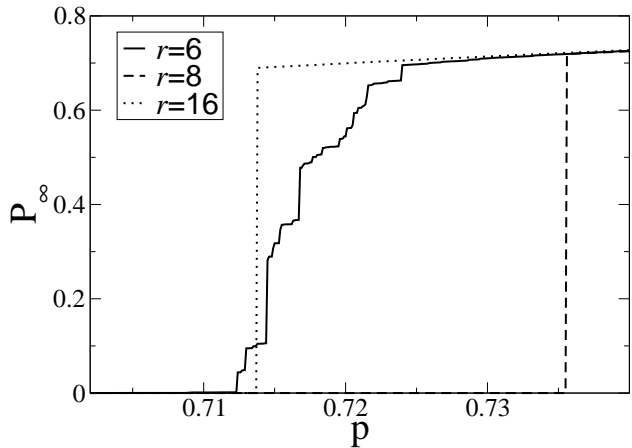}
\caption{The fraction of nodes in the giant component as a function of
  nodes survived after the initial attack. We perform the simulations by
  gradually removing additional nodes. For $r=6$ the decrease of giant
  component occurs in multiple steps, characteristic of a second-order
  transition. For $r=8$ and $r=16$, the giant component may
  completely collapse by removal of even a single additional node,
  characteristic of a first-order transition.
\label{giant8.ps}}
\end{figure}

What can we learn from the flat interface behavior on our original
system with only initial random failures? When $r$ is large, in the absence of
artificial flat interface, the system does not collapse but
rather stays in a metastable state where $p_c^\mu<p<p_c^f$.
As soon as $p=p_c^\mu$, a hole of size $r$ is spontaneously formed
in a low $p$ regime and its interface freely propagates through the
system---because $p$ is already below the critical point $p_c^f$ of the
interface propagation.  As a result, the interface will completely wipe
out the remaining giant component [see
  Fig.~\ref{pictures}(c)--(e)]. Thus for large $r$, the transition is
first order, meaning it is all or nothing, a transition similar to
spontaneous nucleation.  At these conditions, the removal of even a
single additional node may cause the disintegration of the entire system
(Fig.~\ref{giant8.ps}).

\begin{figure}
\includegraphics[width=0.4\textwidth, angle = -90]{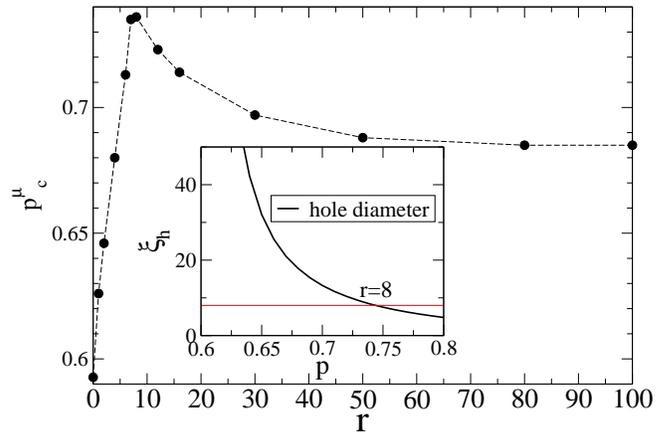}
\caption{The critical $p_c^\mu$ as a function of interdependent distance
  $r$. The change from second to first order transition occurs at
  $r_{\rm max}\approx 8$. The critical $p_c^\mu$ of mutual percolation
  linearly increases for $r<r_{\rm max}$ following the percolation
  threshold for flat interface and then gradually decreases to
  $p_c^\mu=0.683$ at $r= \infty$, which is in good agreement with the
  theoretical results. Inset: Diameter of the hole size $\xi$ as a
  function of $r$ on conventional percolation on a single lattice
  network.  $\xi_h\approx r_{\rm max} = 8$ at $p=0.744$ is in good agreement
  with the simulation.
\label{pcvsr}}
\end{figure}

The dynamics of the system becomes completely different for small $r$.  In
this case, when $p_c^f$ is small, the characteristic size of the holes
$\xi_h$ in the percolation cluster is sufficiently large and there are
many holes of size $\xi_h(p_c^f)>r$. Thus, the flat interface is formed
before it begins to propagate. Once $p$ approaches $p_c^f$ from above,
the interface begins to propagate simultaneously from all large holes in
the system. It can spontaneously stop at any stage of the cascade,
leaving any number of sites in the mutual giant component
(Fig.~\ref{giant8.ps}). The average number of sites in the giant
component will approach zero as $p$ approaches $p_c^f$, subject to
strong finite-size effects as in conventional percolation. So for small
$r$, the transition is a second-order, and
$p_c^\mu(r)=p_c^f$ linearly increases with $p$ (Fig.~\ref{pcvsr}).

The inset of Fig.~\ref{pcvsr} shows that at $r=r_{\rm max}$,
$\xi_h(p_c^f(r))=r\approx 8$, and a flat interface will not
spontaneously form. Thus $p$ must be below $p_c^f(r)$ in order for the
hole of size $r$ to appear in the system. Once a single hole of such
size appears, the flat interface will freely propagate below its
critical threshold wiping out the entire coupled network system,
as in a first-order transition.  Note that
$p_c^f(r_{\rm max})\approx 0.738>p_c^\mu=0.6827$. Thus as $r$ increases,
$p_c^\mu(r)$ gradually decreases (Fig.~\ref{pcvsr}). This gradual
decrease is caused by two factors.  When $r$ increases in the vicinity
of $r_{\rm max}$, smaller and smaller $p$ is needed in order to create holes
of size $r$. When $p$ becomes close to $p_c^\mu$, the system begins to
undergo local cascades of failures if the average density in the region
of size $r$ falls below $p_c^\mu$. The average over $r^2$ nodes of this
region can deviate from the mean $p$ on the order of a standard
deviation $\sqrt{p(1-p)}/r$, thus making the disintegration possible if
$p=p_c^\mu(r)\approx p_c^\mu+C/r$, where $C$ is a constant. Note that
$p_c^\mu(r)$ has a tendency to increase with the system size. The larger
is the system, the more likely a sufficiently large hole or a
sufficiently large fluctuation in local density will lead to a local
cascade of failures.

In summary, our analysis suggests that the change from second-order to
first-order transition occurs at $r_{\rm max}\approx 8$. Note that
Ref.~\cite{Son2011} found a second-order transition for $r=0$ on two
interdependent lattice networks. Our studies show rich phase transition
phenomena when the length of the dependency links $r$ changes.  The
critical $p$ of mutual percolation increases linearly with $r$ in the
range of $r<r_{\rm max}$, and is characterized by a second-order
transition.  For $r\ge r_{\rm max}$, the cascading failures suggest
a first-order transition and the critical
$p$ gradually decreases to $p_c^\mu=0.683$ for $r\to \infty$.

\bigskip

\noindent
We thank an anonymous referee for comments that improved the paper. We
also thank DTRA and ONR for support. SVB acknowledges the partial
support of this research through the Dr. Bernard W. Gamson Computational
Science Center at Yeshiva College. SH thanks the LINC and the Epiwork EU
projects, the DFG, and the Israel Science Foundation for support.

\end{document}